# Cointegration of SARS-CoV-2 Transmission with Weather Conditions and Mobility during the First Year of the COVID-19 Pandemic in the United States.

Hong Qin
Dpt. of Comp. Sci. & Eng.
U. of Tennessee
Chattanooga, TN, U.S.A.
hong-qin@utc.edu
Corresponding author

Syed Tareq
Dpt. of Civil & Chem. Eng.
U. of Tennessee
Chattanooga, TN, U.S.A.
zhd777@mocs.utc.edu

William Torres
Dpt. of Comp. Sci.
Adelphi U.
Garden City, NY, U.S.A.
willtorres@mail.adelphi.edu

Megan Doman
Dpt. of Comp. Sci. & Eng.
U. of Tennessee
Chattanooga, TN, U.S.A
syn427@mocs.utc.edu

Cleo Falvey
Ctr. for Comp. & Integ. Biol.
Rutgers State U. of New Jersey
Camden, NJ, U.S.A.
chf29@scarletmail.rutgers.edu

Jamaree Moore
Dpt of Biology
Norfolk State U.
Norfolk, VA, U.S.A.
j.s.moore97144@spartans.nsu.edu

Meng Hsiu Tsai
Dpt. of Comp. Sci. & Eng.
U. of Tennessee
Chattanooga, TN, U.S.A.
wmf223@mocs.utc.edu

Yingfeng Wang
Dpt. of Comp. Sci. & Eng.
U. of Tennessee
Chattanooga, TN, U.S.A.
yingfeng-wang@utc.edu

Azad Hossain
Dpt of Biol.Geol & Env. Sci.
U. of Tennessee
Chattanooga, TN, U.S.A.
azad-hossain@utc.edu

Mengjun Xie
Dpt. of Comp. Sci. & Eng.
U. of Tennessee
Chattanooga, TN, U.S.A.
mengjun-xie@utc.edu

Li Yang
Dpt. of Comp. Sci. & Eng.
U. of Tennessee
Chattanooga, TN, U.S.A.
li-yang@utc.edu

*Abstract*— **Correlation between weather and the transmission of SARS-CoV-2 may suggest its seasonality. We examined the cointegration of virus transmission with daily temperature, dewpoint, and confounding factors of mobility measurements during the first year of the pandemic in the United States. We examined the cointegration of the effective reproductive rate, Rt, of the virus with the dewpoint at two meters, the temperature at two meters, Apple driving mobility, and Google workplace mobility measurements. Dewpoint and Apple driving mobility are the best factors to cointegrate with Rt. The optimal lag is two days for cointegration between Rt and weather variables, and three days for Rt and mobility. We observed clusters of states that share similar cointegration results, suggesting regional patterns. Our results support the correlation of weather with the spread of SARS-CoV-2 and its potential seasonality.**

## I. INTRODUCTION

Coronavirus infectious diseases are often seasonal and are influenced by weather [1, 2]. The emergence of the SARS-CoV-2 has quickly established its presence in the population, and COVID-19 has been predicted to last for years to come [3]. Therefore, it is of great interest whether the SARS-CoV-2 pandemic would become a seasonal infectious disease. Because aerosols and droplets are major ways of transmission for coronaviruses, the weather is expected to influence their transmission, and low humidity and low temperatures are expected to be associated with increased viral transmission [2, 4, 5].

Standard correlation methods have previously been used to examine the potential relationship between weather conditions and COVID-19 transmission. For example, based on the Pearson correlation method, daily COVID-19 cases in the U.S.A. were weakly correlated with air temperature and social distancing indices[6]. Based on Spearman rank correlation tests, air temperature and other weather conditions correlate with daily COVID-19 cases and deaths in Madrid, Spain [7]. Based on an additive model and piecewise linear regression, ambient temperature correlated with COVID-19 cases in 122 cities in China [8]. A significant Spearman correlation was reported between average daily temperature and reported COVID-19 cases in Jakarta, Indonesia [9]. Based on descriptive statistics, U.S. states with absolute humidity at a lower range tend to have high numbers of daily COVID-19 cases [10].

Standard regression methods can lead to spurious correlations in time series data, and for this reason, the cointegration method was developed to examine the relationship between time series [11]. It is worth emphasizing that COVID-19 daily cases and weather conditions are time series data. Hence, we chose to use cointegration to examine the relationship between COVID-19 transmission, weather conditions, and confounding mobility data. The mobility data, generated on cellphone usage, were designed to describe social distancing practices in communities.

We also chose to use the effective reproductive number, Rt, also known as the time-varying reproduction number to describe the transmission of SARS-CoV-2, instead of the daily cases. The Rt represents the average number of secondary infections caused by each new infection. Estimation of Rt can account for

We thank the support of NSF #1761839, #1852042, #2149956, an internal CEACSE, and the Office of Vice Chancellor of Research at the University of Tennessee at Chattanooga.

reporting delays in the reported cases of COVID-19, daily cases fluctuation, and reporting delays [12].

The initial wave of the pandemic is expected to be heavily influenced by the susceptibilities of the population to a new pathogen and is hence less likely to be informative on weather conditions [1]. Therefore, we excluded the initial wave and started our analysis from May 1, 2020, to February 15, 2021, for 290 days. We stopped our analysis on February 15, 2021, when about 5% of the US population had received the second dose of the COVID-19 vaccine.

## II. MATERIALS AND METHODS

### A. Data Sources

We obtained confirmed cases for each county in the U.S.A. from the COVID-19 Data Repository by the Center for Systems Science and Engineering at Johns Hopkins University [13]. We retrieved the Apple mobility report from its website on November 16, 2021. We retrieved the Google community mobility report from its website on November 17, 2021. We wrote a Python script to parse out the driving mobility for each county in the U.S.A. We found 2638 counties in the Apple mobility report, and 2837 counties in the Google mobility report. We found 2834 counties are shared by the three data sets of COVID-19 cases, Apple and Google mobility reports. Among these counties, many did not have good-quality cases numbers during the 290-day window of our analysis and were later dropped from further analysis (detailed in results and Table 1).

We downloaded weather data from Copernicus Climate Change Service Climate Data Store in GRIB format [14] in November 2021. The ERA5-Land hourly data include 2-meter temperature and 2-meter dew point temperature in Kelvin. The 2-meter temperature (T2m) is the temperature at 2 meters above the surface of the Earth. The 2-meter dew-point temperature is the temperature to which air at 2 meters above the surface of the Earth would have to be cooled for saturation to occur and is a measure of the humidity of the air. Both 2-m temperature and dew-point temperature were estimated by interpolating between the lowest model level and the Earth's surface, taking account of the atmospheric conditions. ERA5-Land data is provided in a latitude-longitude grid of 0.1 degree x 0.1 degree, corresponding to a resolution of about 9 km. The ERA5 data is released up to date three months prior to the date that we retrieved the data. We wrote a Python script to parse weather conditions for specification geographic locations from the ERA5-Land hourly data in GRIB format. For each county, we parsed a 0.6x0.6 degree area for each county with the reported latitude and longitude as the center, and estimated the average measurement at 16:00 on each day from each county. Both T2m and dewpoint are measured in Kelvin.

### B. Time Series Analyses.

We applied the Augmented Dickey-Fuller (ADF) test to the stationarity of time series data, using the *adf.test* function from the R *tseries* package [15]. Time series with an ADF p-value of less than 0.05 are considered stationary data and were excluded from the cointegration test.

We applied the Johansen test to examine the long-term cointegration of two or three non-stationary time series with the *ca.jo* function of R *urca* package [16]. We used a critical value of 0.01 to select the number of cointegration vectors r, and the first non-rejection of the null hypothesis is taken as an estimate of r. We explore the lag from two days to 20 days. The minimum lag of *ca.jo* test function is two days. We exclude the data sets that resulted in numeric errors during the cointegration test, likely due to frequent missing values.

### C. Computing.

It is computationally intensive to extract weather conditions from thousands of locations to cover the period of the pandemic from the downloaded GRIB file. It is also computationally intensive to estimate Rt for thousands of counties in the U.S.A. We ran hundreds of parallel jobs using a high-performance computing cluster, Tennessine-117, at the SimCenter at the University of Tennessee at Chattanooga (UTC). Job management scripts were written to handle the computing jobs. For Johansen cointegration tests at all counties, we ran R scripts at a Lambda Linux workstation provided by the College of Engineering and Computer Science at UTC.

Table 1. Percentages of counties with cointegrated time-series factors.

| Lag | Rt T2m (n=1642) | Rt Dewpoint (n=175) | Rt A-driving (n=967) | Rt G-workplace (n=1367) | Rt, T2m A-driving (n=853) | Rt Dewpoint A-driving (n=131) | Rt T2m G-workplace (n=1179) | Rt Dewpoint G-workplace (n=136) |
|---|---|---|---|---|---|---|---|---|
| 2 | **65.4%** | **93.1%** | 72.2% | 49.9% | **39.3%** | **51.9%** | **28.2%** | 39% |
| 3 | 27.9% | 65.7% | **83.7%** | **61.9%** | 13.4% | 35.9% | 17.8% | **42.6%** |
| 4 | 7.4% | 30.3% | 70.4% | 57.7% | 8.0% | 18.3% | 6.8% | 25.0% |
| 5 | 2.2% | 12% | 43.0% | 41.4% | 3.9% | 14.5% | 3.7% | 10.3% |
| 6 | 0.1% | 1.1% | 10.3% | 26.2% | 1.5% | 3.1% | 0.6% | 3.7% |

Results of the cointegration test are presented in columns. Rows indicate different values of the lag parameter in the Johansen test. Inside of the parentheses in the first row of each column, n represents the total number of counties with available nonstationary data sets. The highest percentages of cointegrated counties in each test are highlighted in bold. For two-factor tests, a cointegration rank of 2 was chosen at the critical value of 0.01 for the Johansen test. For three-factor tests, a rank of 3 was chosen.

## III. RESULTS

Below, we will first illustrate the cointegration analyses using two examples, then present the results at the county level, and finally, show the spatial patterns at the state level.

### A. Cointegration can Reveal the Potential Relationship between Virus Transmission, Weather, and Mobility Variables.

The effective reproductive rate, weather conditions, and daily mobility reports are all time series. For time series, standard regression methods tend to lead to spurious correlations. Cointegration commonly occurs when a linear combination of several nonstationary time series variables results in a stationary signal whose mean and variance do not change over time. We used data sets from two counties as examples to illustrate the principle of cointegration analysis (Figure 1A and 1B). In both counties, we can see major peaks and valleys for Rt, dewpoint, and Apple driving mobility variables (The first three sub-figures in Figure 1A and 1B). The cointegrated signals are linear combinations of the three time series variables based on the Eigenvector of the Johansen test (the last sub-figure in Figure 1A and 1B). The cointegrated signals become relatively flat with fluctuating noises over time, and can be considered as stationary based on ADF test (p-value =0.04 and 0.01 for McKenzie, ND and Clay, MN, respectively).

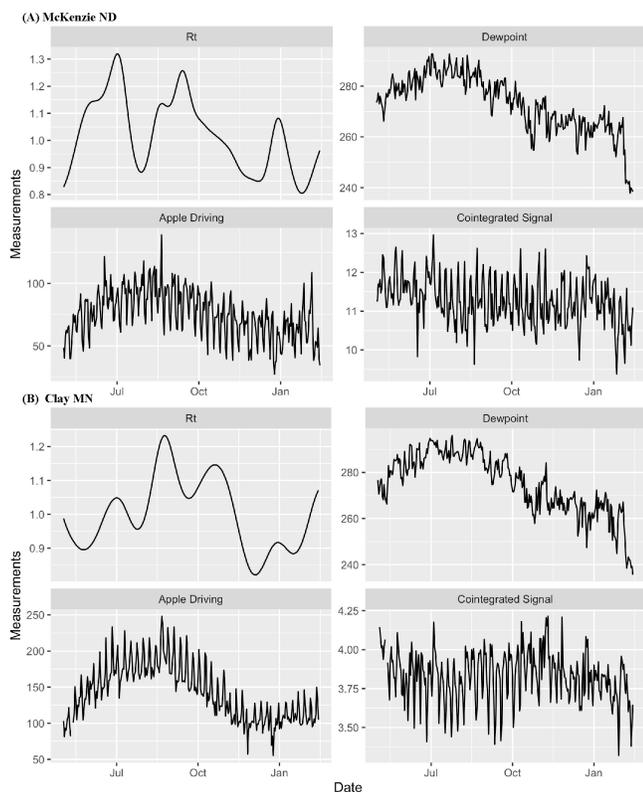

Figure 1. Illustration of cointegration analyses in two example counties. In both counties, three nonstationary time series, Rt, dewpoint, Apple driving mobility cointegrated into a stationary signal in the 4$^{th}$ sub-figures of both counties

### B. Weather and Mobility Variables Cointegrate with the Effective Reproductive Rate at the County Level.

We examined the cointegration of the effective reproductive rate Rt to weather variables (T2m and dewpoint), mobility variables (Apple driving and Google workplace mobility estimates) at all available counties in the U.S.A (Table 1). Because cointegration tests are only valid for non-stationary input, all of the time series data were first tested for stationarity. We filtered out counties in which the input factors are stationary based on ADF test with a p-value less than 0.05. We studied the effect of the lag parameter for the Johansen cointegration test from two days to 20 days. The two-day lag is the minimum allowed lag in the Johansen test. We did not observe any qualitative changes in results when the lag parameter is longer than 6 days.

To cointegrate with Rt, we found that the optimal lag is 2 days for T2m and dewpoint, and 3 days for Apple driving and Google workplace mobility (Table 1). We found that dewpoint cointegrated more frequently with Rt than T2m, with a cautionary note that the 175 nonstationary dewpoint data set was much smaller than the 1642 nonstationary T2m data set. We observed that Apple driving mobility cointegrates with Rt more frequently than Google workplace mobility. The three-factor cointegration results also support that dewpoint and Apple driving mobility are more informative on Rt than the other two variables.

### C. Regional Patterns of Cointegration Results at the State Level.

To examine the regional patterns in cointegration of weather and mobility to Rt, we visualized the percentages of cointegrated counties in each state (Figure 2 and 3). The optimal lag parameters based on these state-level visualization are consistent with the overall analysis in Table 1.

Clusters of states with similar percentages of cointegrated counties can be observed in the Rt and T2m test, and in the Rt and Apple driving mobility test (first and second column of Figure 2). Clustering of states with similar effects are not obvious in the Rt and Google workplace test, and are inconclusive in the Rt and dewpoint test because many states do have available nonstationary dewpoint data sets. Clusters of states with similar effects are more conspicuous in the three-factor cointegration tests (Figure 2). Overall, the regional patterns for the cointegration of temperature to Rt are more conspicuous than that of dewpoint. Both Apple driving and Google workplace mobility generally share comparable regional patterns for cointegration with Rt, but there is clear difference between these mobility proxies in some states. For example, in Alaska, Google workplace mobility cointegrate with Rt better than Apple driving mobility, in the sense that Alaska appears to be redder in the fourth column than the corresponding cells in the third column in Figure 2.

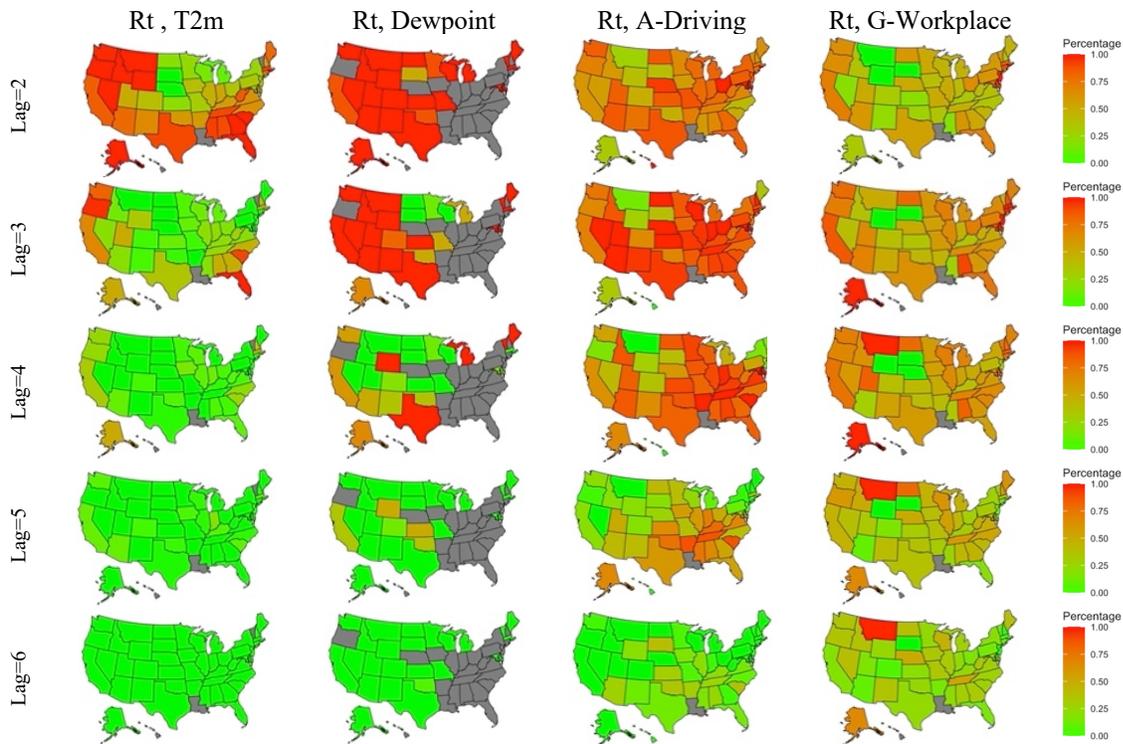

Figure 2. Percentage of counties in which the effective reproductive rate (Rt) cointegrates with the temperature at two meters (T2m), dewpoint, Google workplace mobility (G-workplace), and Apple driving mobility (A-driving) in the US with the lag parameter ranging from two days to six days. Each state is colored based on the percentages of counties with cointegrated two factors as described in the first row. Cointegration of Rt and T2m generally has a lag of 2 days. Cointegration of Rt and dewpoint generally has an optimal lag of 2 days, but is also substantial at a lag of 3 days. Cointegrations of Rt with the two mobility estimates have an optimal lag of 3 or 4 days. The critical value of cointegration tests was chosen at 1%. Augmented Dickey-Fuller (ADF) tests were first applied to all three factors. States in deep red indicate high percentages of counties with cointegrated factors, and states in light green indicate low percentages. Factors were treated as stationary if their ADF test p-values are less than 0.05 and were excluded from the Johansen cointegration test. States with no available non-stationary factors are colored in gray in each column.

## IV. DISCUSSION AND CONCLUSION

Overall, our research detected strong cointegration of Rt, mobility and dewpoint with two or three days of lag. However, we are aware of the limitation of the cointegration method. Because cointegration tests require non-stationary time series, we have to exclude many data sets that could not pass the ADF stationarity test. The 290-day window of the present study maybe not long enough to address long-term cointegration effect. However, longer time windows for studying pandemics in the U.S. have additional complexities. During the second year of the pandemics in the USA, vaccination rates in the population were likely to have a significant role in reducing the spread of the virus [17]. In addition, new variants, such as Delta and Omicron, are known to cause new peaks of daily cases [18]. There is also evidence that mobility data become noise and unreliable to proxy social distancing practice as the pandemic went on [19]. Therefore, our choice of the 290-day enabled us to focus on the roles of a relatively small set of confounding factors, especially for weather and mobility variables.

Dewpoint is a proxy of humidity. Interestingly, our results suggest that humidity cointegrates with Rt better than temperature. It is expected SARS-CoV-2 becomes more transmissible in drier air. It is expected most SARS-CoV-2 transmission occurs indoors. Indoor temperatures are often controlled in the U.S., however, indoor humidity is generally less controlled in most households in the U.S. In one study, indoor and outdoor absolute humidity are highly correlated in Greater Boston, Massachusetts [20].

Interestingly, we found dewpoint and temperature cointegrate with Rt as effectively as mobility measurements. It was argued that weather conditions have much less effect on virus transmission than human mobility and social distancing [21]. It is plausible that weather and human behavior are correlated, in the sense that people tend to stay indoors in cold weather, during which period low humidity associated with winter acted synergistically to drive up viral transmission [22].

We are aware that the cointegration relationship does not necessarily lead to comparable importance in the real world. Our model is a simplification of a complex pandemic, studying a subset of factors that influence the spread of novel pathogens. Other factors have already been shown to impact Rt, but were not included, such as vaccination masking protocols, effective contact tracing, and the quarantine of infected individuals [22]. Therefore, future studies will be needed to disentangle the roles of weather conditions and mobility on viral transmission. Future studies would be needed to examine the roles of weather conditions and mobility on virus transmission.

Overall, we concluded that temperature and dewpoint cointegrated with effective reproductive rate SARS-CoV-2 at comparable levels to those of mobility measurements at the county level in the first year of the pandemic in the U.S.A. Our results align with previous literature that suggests that COVID-19 spread is seasonal and that social distancing measures can curb the spread of novel pandemics. Therefore, our research adds to a growing body of literature that can inform policy decisions and reduce the exponential spread of novel diseases [2].

ACKNOWLEDGMENT

We thank the helpful comments and discussion from Landen Bauder, Derek Campbell, and many faculty and students who participated in the iCompBio programs. We thank the computing support from the SimCenter and the College of Engineering and Computer Science at the University of Tennessee at Chattanooga.

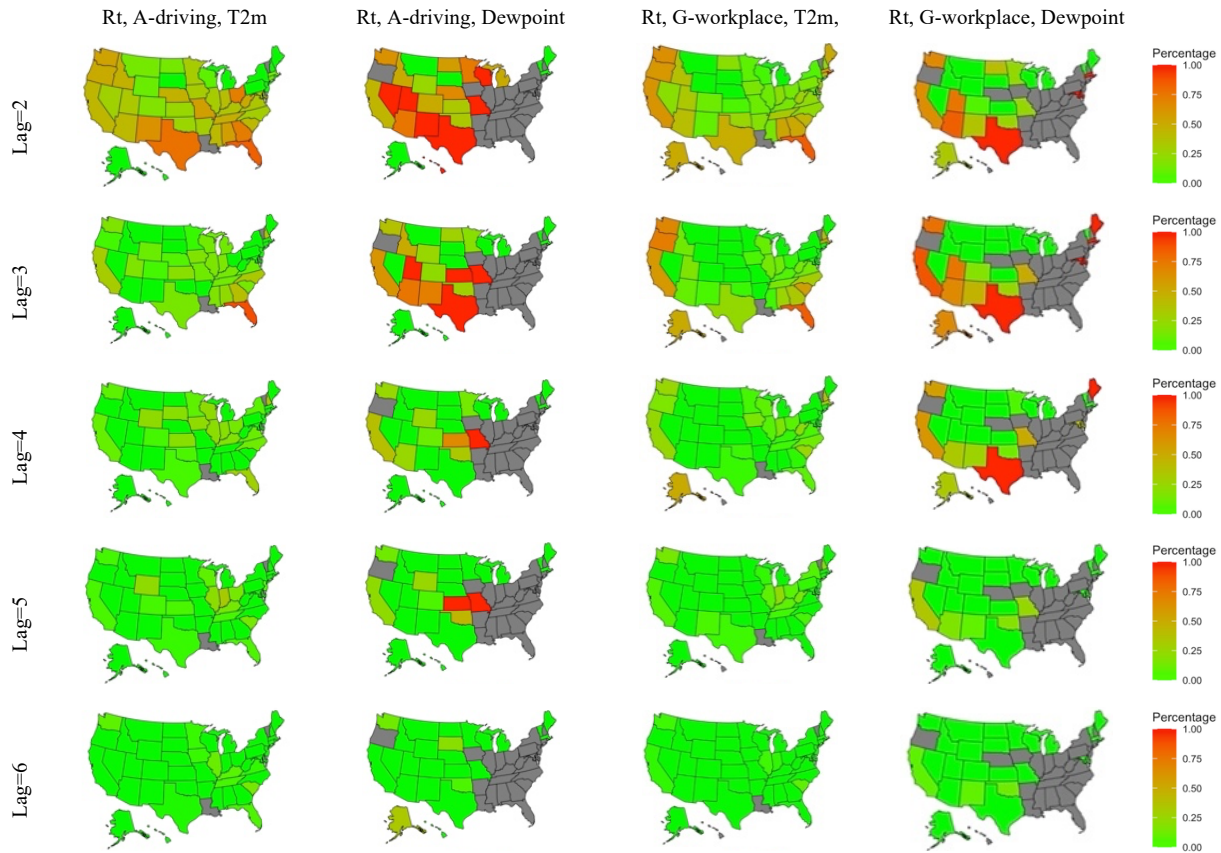

Figure 3. Regional patterns can be observed in three-factor cointegration results at the state level. The best cointegrated three factors are effective reproductive rate (Rt), Apple driving mobility (A-driving), and dewpoint at two meters, with an optimal lag of 2 days. Each state is colored based on its percentage of counties in which three factors are co-integrated. Each column represents a cointegration of the three factors in the first row. Procedures are similar to those in Figure 2.

.